\documentclass[epj]{webofc}
\usepackage[varg]{txfonts}   
\graphicspath{{/home/kuba/DATA/konferencje_seminaria/figs/}}
\woctitle{Physics Opportunities at an Electron-Ion Collider}

\begin{document}
\title{GPDs in heavy meson production and Compton scattering}
%
%
\author{D.Yu. Ivanov\inst{1} \and
        B. Pire\inst{2} \and
        L. Szymanowski\inst{3}\and
        J. Wagner \inst{3}
        }

\institute{Sobolev Institute of Mathematics and Novosibirsk State University, 630090 Novosibirsk, Russia
\and
           Centre de physique th\'eorique,  \'Ecole Polytechnique, CNRS, Universit\'e Paris-Saclay, 91128 Palaiseau, France
           \and
           National Centre for Nuclear Research (NCBJ), Warsaw, Poland
          }

\abstract{%
  Exclusive processes of heavy meson production and spacelike and timelike deeply virtual Compton scattering allow us to investigate the hadron structure in terms of Generalized Parton Distributions (GPDs). We  review recent developments in the NLO description of such processes.
}
\maketitle
\section{DVCS \& TCS}

\begin{figure}[h]
\begin{center}
\includegraphics[keepaspectratio,width=0.4\textwidth,angle=0]{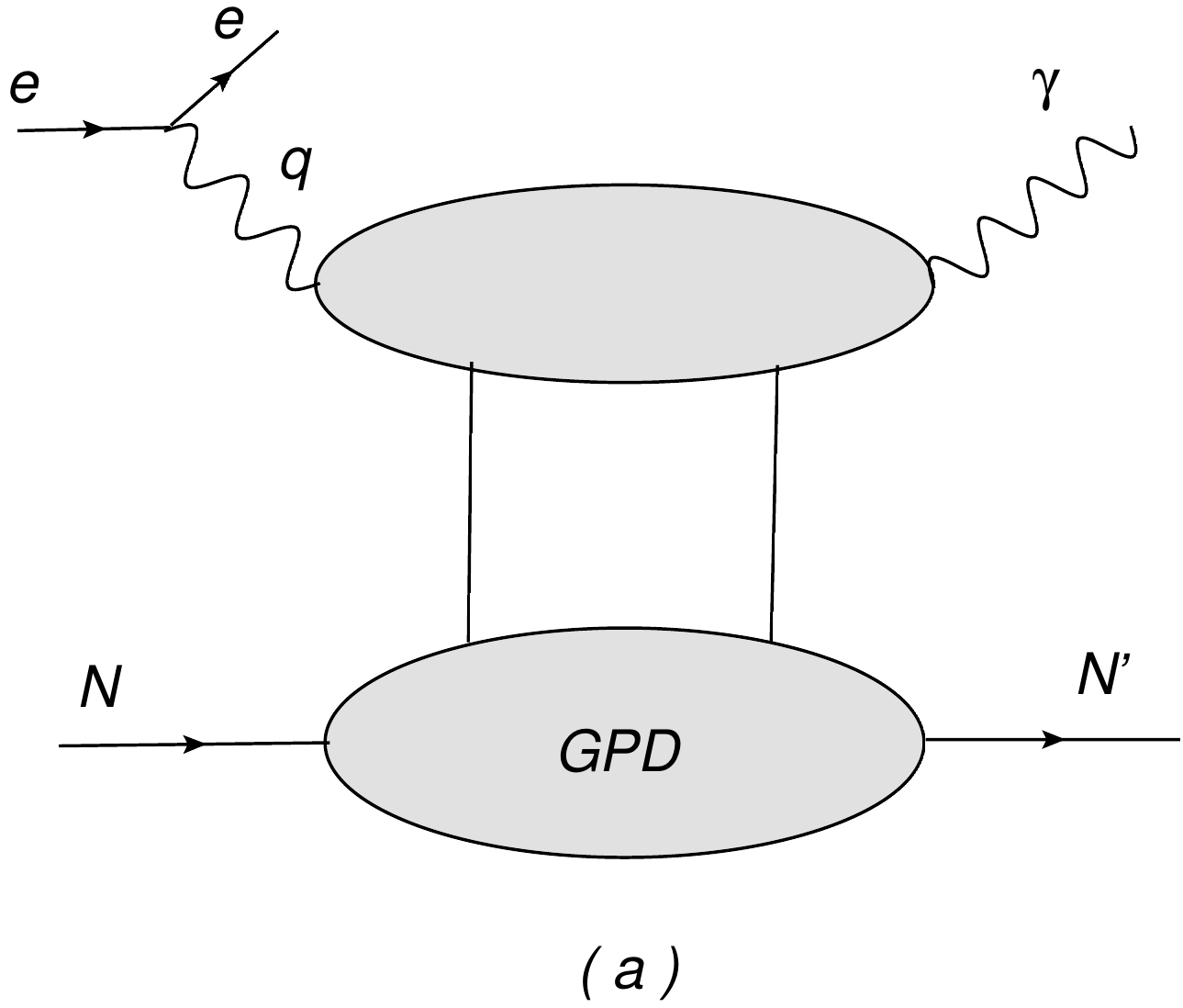}
\includegraphics[keepaspectratio,width=0.45\textwidth,angle=0]{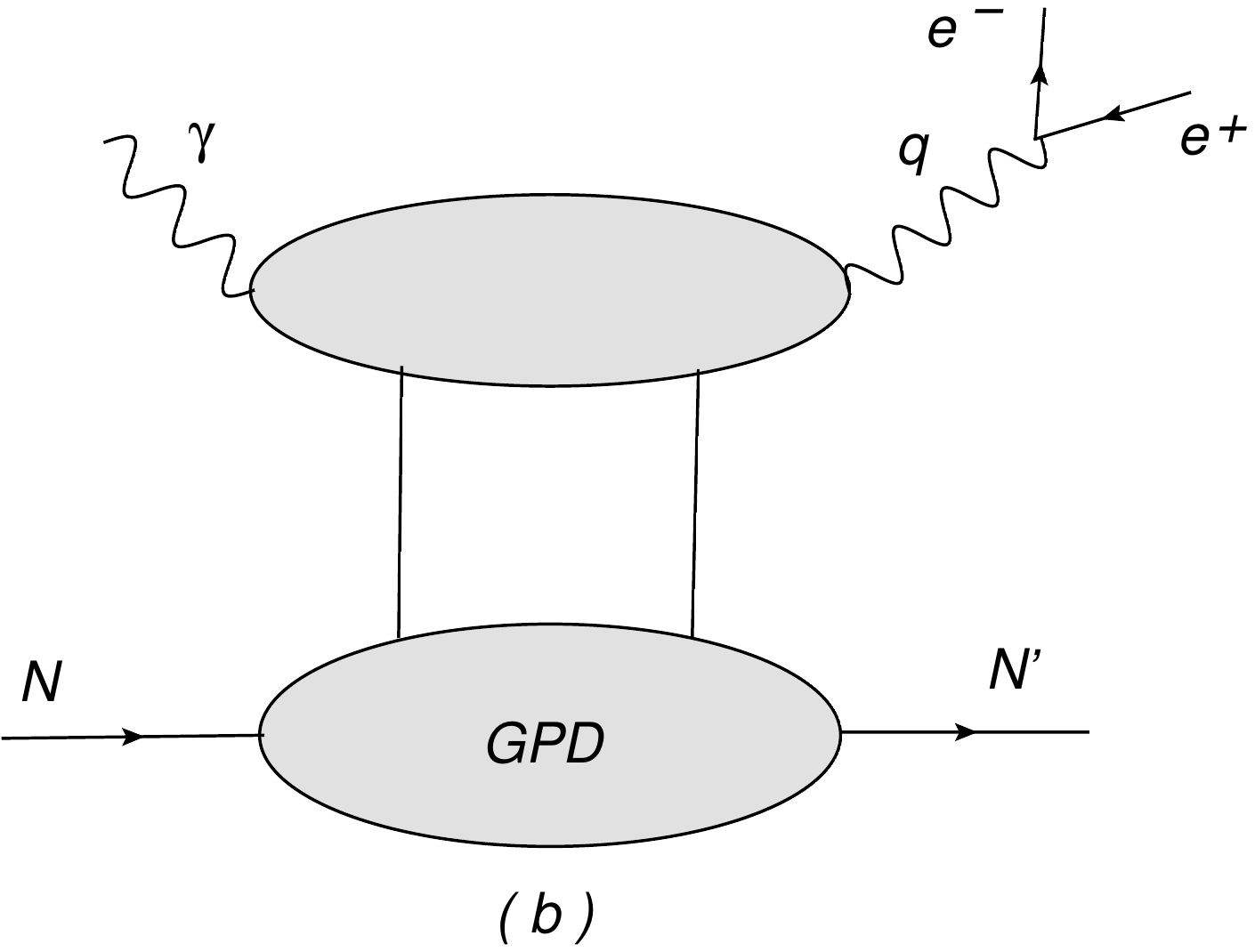}
\end{center}
\label{proc}
\caption{(left)Deeply Virtual Compton Scattering (DVCS) : $l N \to l'N'\gamma $ (right) Timelike Compton Scattering (TCS): $\gamma N \to l^+l^- N'$}
\end{figure}

The studies of deeply virtual Compton scattering (DVCS)  $\gamma^*(Q^2)p\to \gamma p$ in electron-proton collisions
at JLAB, HERA are the primary source of our knowledge on GPDs \cite{gpdrev}. The related  timelike Compton scattering (TCS) process \cite{TCS} shares all its virtues for accessing correlated information on the light cone momentum fraction and the transverse location of partons in hadrons  \cite{Burk}. Both reactions, illustrated in Fig.\ref{proc}, can be seen as limiting cases of the double virtual Compton scattering process,
\begin{equation}
\gamma^*(q_{in}) N (p) \to \gamma^*(q_{out}) N'(p')\, .
\end{equation}
The relevant light-cone ratios describing the processes of interest in the generalized Bjorken limit are the scaling variable $\xi$ and  skewness $\eta  > 0$:
\begin{equation}
\xi = -\frac{q^2_{out}+q^2_{in}}{q^2_{out}-q^2_{in}} \eta\,, \quad
\eta =\frac{q^2_{out}-q^2_{in}}{(p+p')\cdot(q_{in}+q_{out})}\,. 
\label{eq:skewnessdef}
\end{equation}
The scattering amplitude is written in a factorized form as :
\begin{eqnarray}
\mathcal{A}^{\mu\nu}(\xi,\eta,t) = - e^2 \frac{1}{(P+P')^+}\, \bar{u}(P^{\prime}) 
\Bigg[\,
   g_T^{\mu\nu} \, \Big(
      {\mathcal{H}(\xi,\eta,t)} \, \gamma^+ +
     {\mathcal{E}(\xi,\eta,t)} \, \frac{i \sigma^{+\rho}\Delta_{\rho}}{2 M}
   \Big) \nonumber\\  \phantom{AAAAAAAAAAAAAaa}
   +i\epsilon_T^{\mu\nu}\, \Big(
    {\widetilde{\mathcal{H}}(\xi,\eta,t)} \, \gamma^+\gamma_5 +
    {\widetilde{\mathcal{E}}(\xi,\eta,t)} \, \frac{\Delta^{+}\gamma_5}{2 M}
    \Big)
\,\Bigg] u(P) \, , 
\label{eq:amplCFF}
\end{eqnarray}
where the Compton form factors are defined as :
\begin{eqnarray}
\mathcal{H}(\xi,\eta,t) &=& + \int_{-1}^1 dx \,
\left(\sum_q T^q(x,\xi,\eta)H^q(x,\eta,t)
 + T^g(x,\xi,\eta)H^g(x,\eta,t)\right)  \,,\\
\widetilde {\mathcal{H}}(\xi,\eta,t)&=& - \int_{-1}^1 dx \,
\left(\sum_q \widetilde{T}^q(x,\xi,\eta)\widetilde{H}^q(x,\eta,t) 
+\widetilde T^g(x,\xi,\eta)\widetilde{H}^g(x,\eta,t)\right)\,,\nonumber
\label{eq:CFF}
\end{eqnarray}
and similarly for ${\mathcal{E}}(\xi,\eta,t)$ and ${\widetilde{\mathcal{E}}(\xi,\eta,t)}$.

The study of $O(\alpha_s$) corrections to the DVCS and TCS amplitudes turns out to be full of surprises. 
Without entering a detailed analysis \cite{PSW} the renormalized coefficient functions for DVCS are given by
\begin{eqnarray}
T^q(x)&=& \left[ C_{0}^q(x) +C_1^q(x) + {\ln\left(\frac{Q^2}{\mu^2_F}\right)} \cdot C_{coll}^q(x)\right] - ( x \to -x )  \,,\nonumber\\
T^g(x) &=& \left[ C_1^g(x) +{\ln\left(\frac{Q^2}{\mu^2_F}\right)} \cdot C_{coll}^g(x)\right] +( x \to -x )\,,\nonumber\\
\widetilde{T}^q(x)&=& \left[
\widetilde{C}_{0}^q(x) +\widetilde{C}_1^q(x) +{\ln\left(\frac{Q^2}{\mu^2_F}\right)} \cdot \widetilde{C}_{coll}^q (x)\right]
+( x \to -x )
\,,\nonumber\\
\widetilde{T}^g(x) &=&   \left[
\widetilde{C}_1^g(x) +{\ln\left(\frac{Q^2}{\mu^2_F}\right)} \cdot \widetilde{C}_{coll}^g(x)\right] - ( x \to -x )\,. \nonumber
\label{eq:ceofficients}
\end{eqnarray} 

\begin{figure}[h]
\begin{center}
\includegraphics[keepaspectratio,width=0.9\textwidth,angle=0]{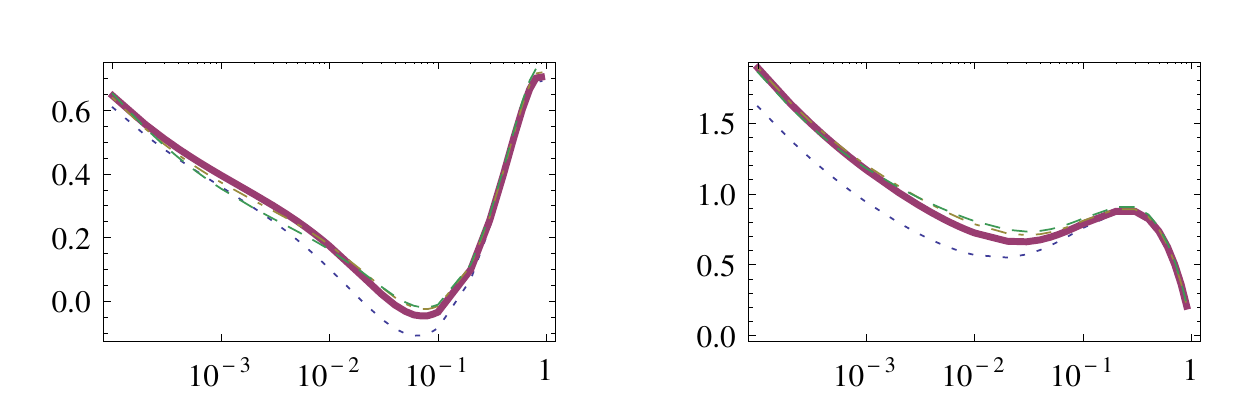}\\
\caption{Full NLO result for DVCS Compton Form Factors, as a function of $\xi$. Left column - $\xi\cdot Re(\mathcal{H}(\xi))$, right column - $\xi\cdot Im(\mathcal{H}(\xi))$, $Q^2 = 4 GeV^2$, $\mu_F^2 = Q^2, Q^2/2, Q^2/3, Q^2/4$ (dotted, solid, dash-dotted, dashed lines respectively).}
\label{NLO1}
\end{center}
\end{figure}

\begin{figure}[h]
\begin{center}
\includegraphics[keepaspectratio,width=1.\textwidth,angle=0]{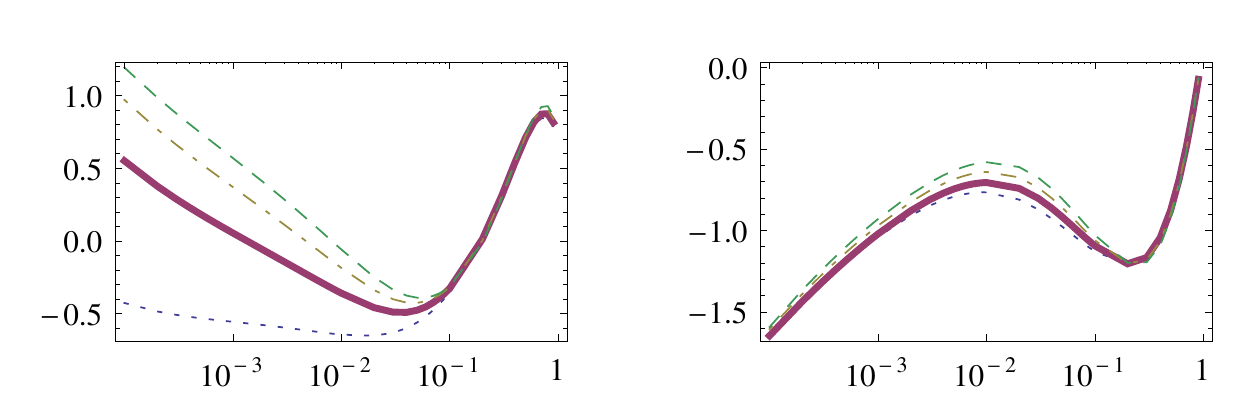}\\
\caption{Full NLO result for TCS Compton Form Factors, as a function of $\xi$. Left column - $\xi\cdot Re(\mathcal{H}(\xi))$, right column - $\xi\cdot Im(\mathcal{H}(\xi))$, $Q^2 = 4 ~GeV^2$, $\mu_F^2 = Q^2, Q^2/2, Q^2/3, Q^2/4$ (dotted, solid, dash-dotted, dashed lines respectively). }
\label{NLO2}
\end{center}
\end{figure}

 Thanks to the analytic structure (in $Q^2$) of the amplitude, the results for {DVCS} and {TCS} cases are simply related \cite{MPSW}: 
\begin{eqnarray}
^{TCS}T(x,\eta) = \pm \left(^{DVCS}T(x,\xi=\eta) +  { i \pi} \cdot C_{coll}(x,\xi = \eta)\right)^* \,, \nonumber
\label{eq:TCSvsDVCS}
\end{eqnarray}
where $+$~$(-)$ sign corresponds to vector (axial) case. This difference has very important phenomenological consequences \cite{MPSSW}. $O(\alpha_s$) corrections are not small and one may question the relevance of phenomenological studies based on $O(\alpha_s$) coefficient functions. Moreover, the factorization scale dependence turns out to be rather large, as can be seen on Fig. \ref{NLO1} and Fig. \ref{NLO2}. Resumming higher order corrections may help to stabilize this unwanted feature. In any case, this is a needed improvement of the theoretical description of DVCS and TCS (see for instance \cite{Altinoluk:2012nt}),  in particular for the high energy domain which is already accessible thanks to the ultra peripheral reactions in hadron colliders \cite{TCSUPC}. This will be a central issue for the phenomenology of the electron-ion collider \cite{EIC}.

\section{Heavy vector meson production}
\label{sec-1}
The photoproduction of a heavy vector meson:
\begin{equation}
\gamma p \to V p
\label{VMprocess}
\end{equation}
is a subject of intense experimental \cite{VMexp} and theoretical \cite{VMth} studies. One motivation of such studies is the possibility to explore gluon GPDs in the nucleon. We present here preliminary results \cite{Prep} on the use of the collinear factorization approach at the next to leading order in $\alpha_s$, which was developed in \cite{HVMP} , in the context of ultraperipheral collisions.

The amplitude $\mathcal{M}$ of the process (\ref{VMprocess})  is given by the factorization formula:
\begin{eqnarray}
{\cal M}
&\sim &\left(\frac{\langle O_1 \rangle_V}{m^3}\right)^{1/2}
 \int\limits^1_{-1} dx
\left[\, T_g( x,\xi)\, F^g(x,\xi,t)+
T_q (x,\xi) \sum_{q=u,d,s}  F^q (x,\xi,t) \, 
\right] \, ,
\end{eqnarray}
with
${F}^{g(q)} (x,\xi,t;\mu_F^2)$  the gluon (quark)  GPDs;  $m$ is a pole mass of the heavy quark, and
$\xi=M^2/(2W^2-M^2)$ is the skewness parameter.

\begin{figure}[h]
\begin{center}
\includegraphics[keepaspectratio,width=.4\textwidth,angle=0]{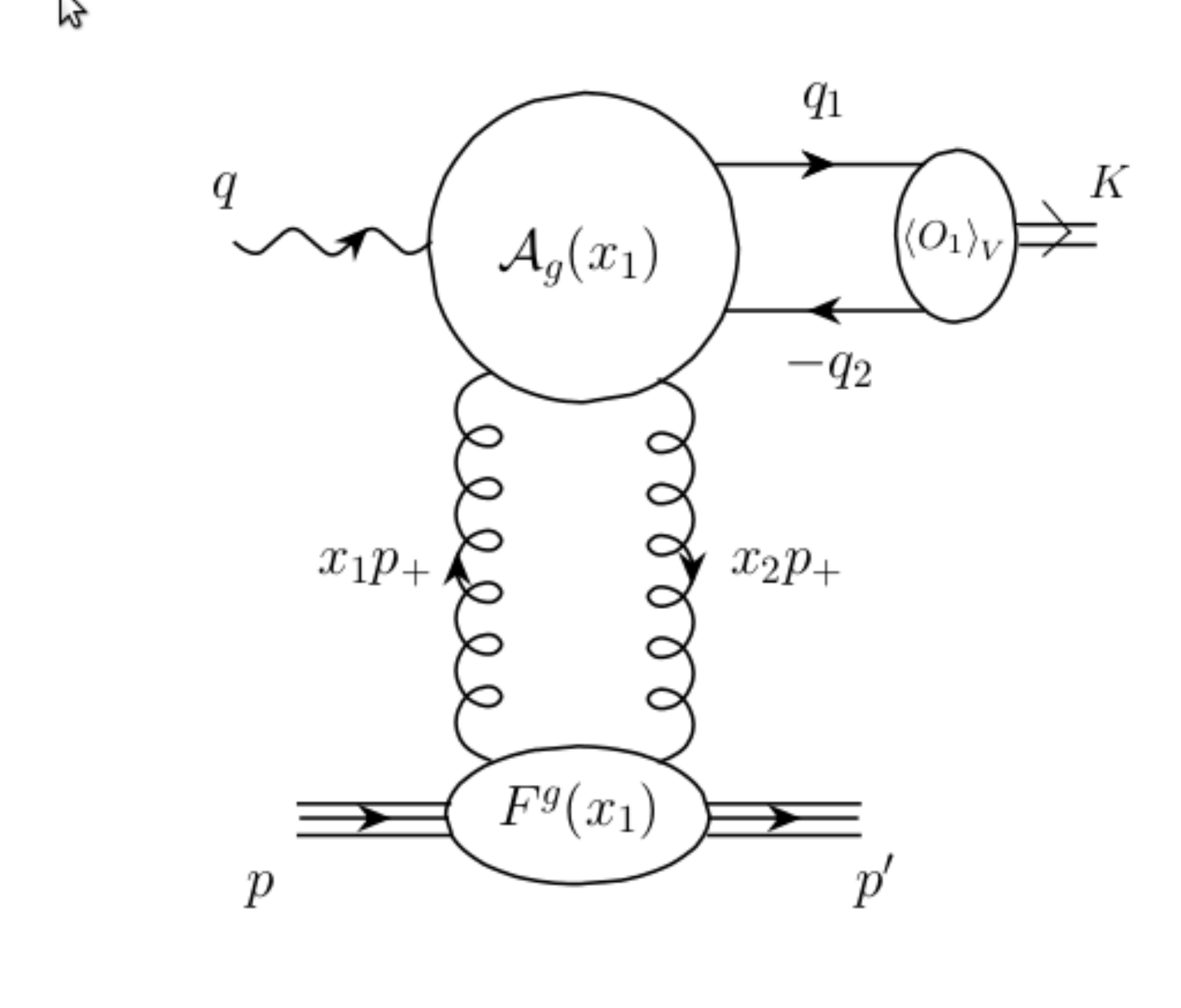}\\
\caption{Kinematics of the heavy vector meson photoproduction}
\end{center}
\end{figure}
All information about the quarkonium structure is encoded
in the NRQCD \cite{Bodwin:1994jh} matrix element $\langle O_1 \rangle_V$ which enters the leptonic decay
rate
$$
\Gamma[V\to l^+l^-]=\frac{2e_q^2\pi\alpha^2}{3}
\frac{\langle O_1\rangle_V }{m^2}
\left( 1-\frac{8\alpha_S}{3\pi}\right)^2 .
$$
The coefficient functions read
\begin{eqnarray}
&&
 T_g(x,\xi)=\frac{\xi}{(x-\xi+i\varepsilon)(x+\xi-i\varepsilon)}
{\cal A}_g\left(\frac{x-\xi+i\varepsilon}{2\xi}\right) \, ,
\nonumber \\
	&&
T_q( x,\xi)={\cal A}_q\left(\frac{x-\xi+i\varepsilon}{2\xi}\right) \, .\nonumber
\label{gAT}
\end{eqnarray}
At leading order
$
{\cal A}_g^{(0)}(y)=\alpha_S \, , 
{\cal A}_q^{(0)}(y)=0 \, .$
The inclusion of NLO corrections has dramatic effects on the production cross section :  NLO corrections are very big and the overall result depends very strongly on the choice of the factorization scale, especially  for the high values of $W$.\footnote{In our calculations, in both LO and NLO cases, we keep the value of renormalization scale fixed $\mu_R=M_{J\psi}$.}

Why are NLO corrections so large in this case, where  $\xi\ll 1$?  The inspection of
NLO hard-scattering amplitudes shows that the imaginary part of the amplitude dominates and
that the leading contribution to the NLO correction originates from the broad integration region
 $\xi\ll x\ll 1$ , where the gluonic part approximates ($\bar \alpha_s=3\alpha_s/\pi$):
\begin{equation}
Im {\cal M}^g\sim H^g(\xi,\xi)+\bar \alpha_s \left[\log\frac{M_V^2}{\mu_F^2}-\log 4\right]
\int\limits^1_{\xi}\frac{dx}{x} H^g(x,\xi)\, .
\label{C1}
\end{equation}
Given the behavior of the gluon GPD at small $x$,
$H^g(x,\xi)\sim xg(x)\sim \rm{const}$, we see that the NLO correction is parametrically large, $\sim \ln(1/\xi)$, and negative unless one chooses the value of the factorization scale sufficiently lower than the hard kinematic scale, $Q=M_V$. 

The size of the corrections, and the sensitivity of the NLO result to the factorization scale choice, shows that some additional information about still higher order contributions  is needed to provide reliable theoretical predictions.
This may come from some strategy for the scale choice to minimalize the one-loop corrections . In our opinion the most promising approach is related with the resummation of the higher orders terms enhanced at small $\xi$ by  powers of large
logarithms of energy, $\sim \bar \alpha_s^n \ln^n (1/\xi)$, see
\cite{Dima:Blois}:
\begin{equation}
{\cal I}m {\cal M}^g\sim H^g(\xi,\xi)
+ \int\limits^1_{\xi}\frac{d x}{x} H^g(x,\xi)
\sum\limits_{n=1}C_n(L)\frac{\bar \alpha_s^n}{(n-1)!}\log^{n-1}\frac{x}{\xi}\, ,
\end{equation}
where
$C_n(L)$ are polynomials of $L=\ln(M_V^2/\mu_F^2)$ which maximum power is $L^n$.
For DIS inclusive structure functions $F_T$ and $F_L$ corresponding $C_n(L)$ coefficients were calculated long time ago
by Catani and Hautman \cite{Catani:1994sq}. Their method developed for  inclusive DIS can be
straightforwardly generalized to exclusive, nonforward processes.

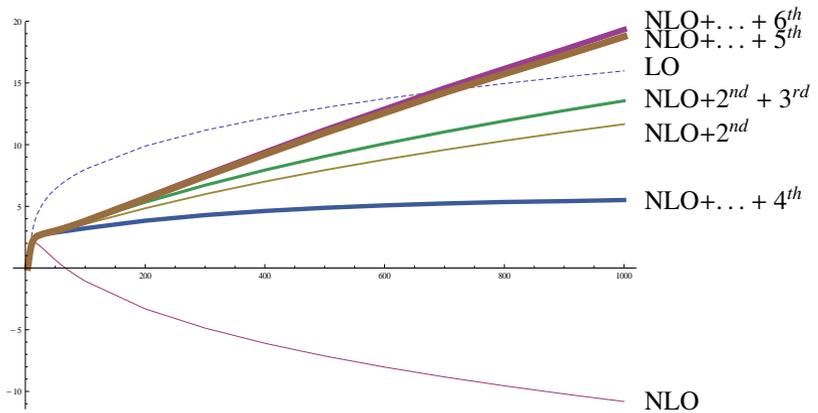
\begin{figure}[h]
\begin{center}
\begin{picture}(240,150)
\put(0,0){\includegraphics[scale=.5]{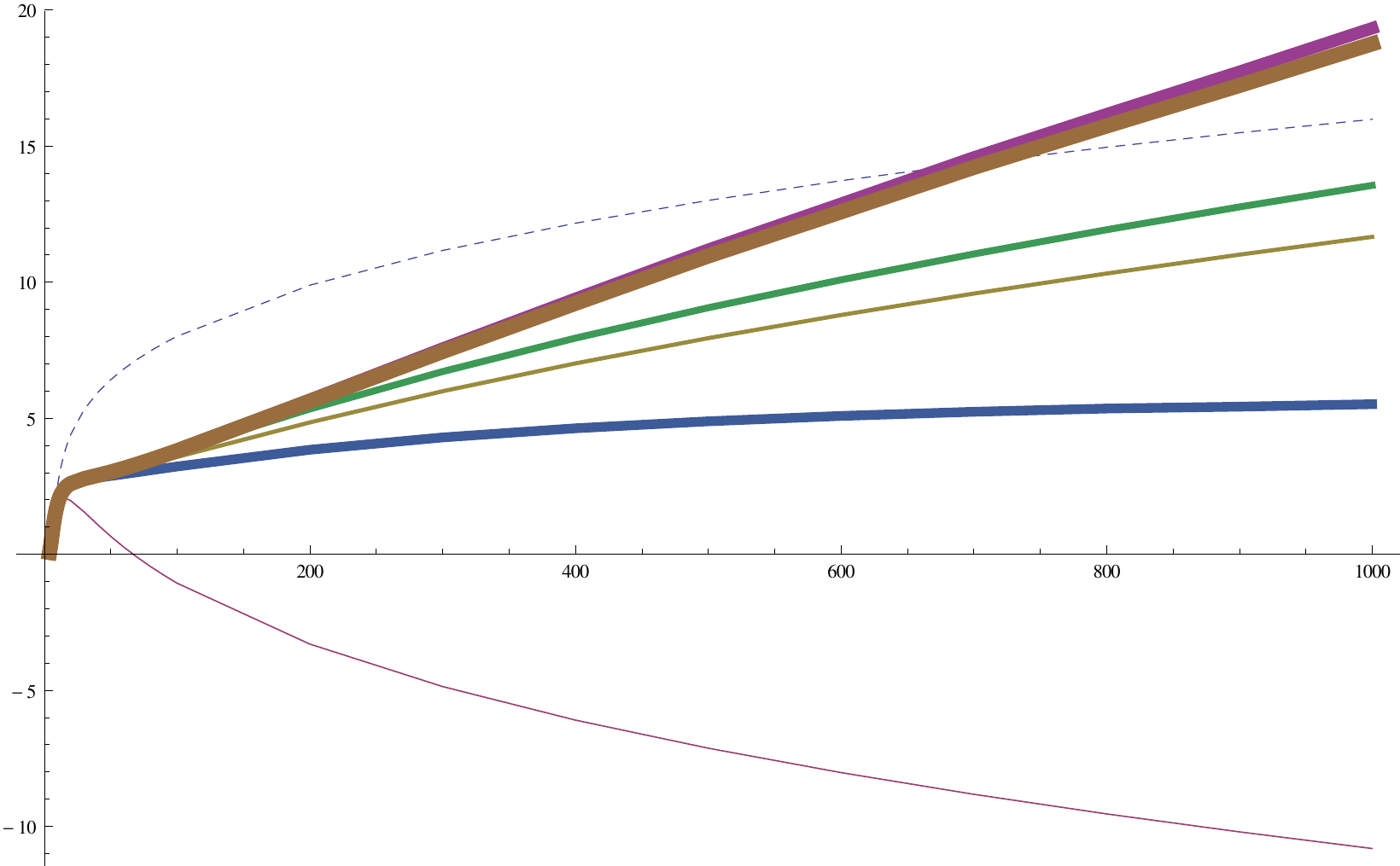}}
\put(240,00){NLO	}
\put(240,125){LO}
\put(240,100){NLO+$2^{nd}$}
\put(240,113){NLO+$2^{nd}+3^{rd}$}
\put(240,75){NLO+$\ldots+4^{th}$}
\put(240,135){NLO+$\ldots+5^{th}$} 
\put(240,142){NLO+$\ldots+6^{th} $}
\end{picture}
\end{center}
\caption{Resummation of the (gluonic GPDs dependent) imaginary part of amplitude (normalized in such way that the LO result equals $H^g(\xi,\xi,t,\mu_F)$) up to the sixth higher order term, as a function of $W$, for $\mu_F=M_V$.}
\label{fig:resum}
\end{figure}

Resummed coefficient functions, parameterized by  $C_n(L)$ polynomials,  can be calculated and conveniently represented
using Mellin transformation. In the Mellin space the resummed coefficient function
is a polynomial in the variable $z=\bar \alpha_s/N$. Conversely, the contributions proportional to the $n^{th}$ power of this variable 
generate terms $\sim \bar \alpha_s^n \ln^n (1/\xi)$ in the process amplitude.
For meson photoproduction our result in the  $\overline{MS}$ scheme reads
$$
1+
z (L-\ln 4)+
\frac{z^2}{6}\left(\pi^2+3\ln^2 4+3L(L-\ln 16)\right)+
 \dots \, ,
$$
where only two nontrivial terms of the expansion are shown.
This leads to $C_1(L)=L-\ln 4$, in accordance with the found high energy asymptotic of NLO result in Eq. (\ref{C1}),
$C_2(L)=\left(\pi^2+3\ln^2 4+3L(L-\ln 16)\right)/6$,
and so on. On the Fig.\ref{fig:resum} we present the effect of such resummation of the imaginary part of amplitude (normalized in such way that the LO result equals $H^g(\xi,\xi,t,\mu_F)$), as a function of $W$ for $\mu_F = M_V$.

\begin{figure}[h]
\begin{center}
\includegraphics[scale=.4]{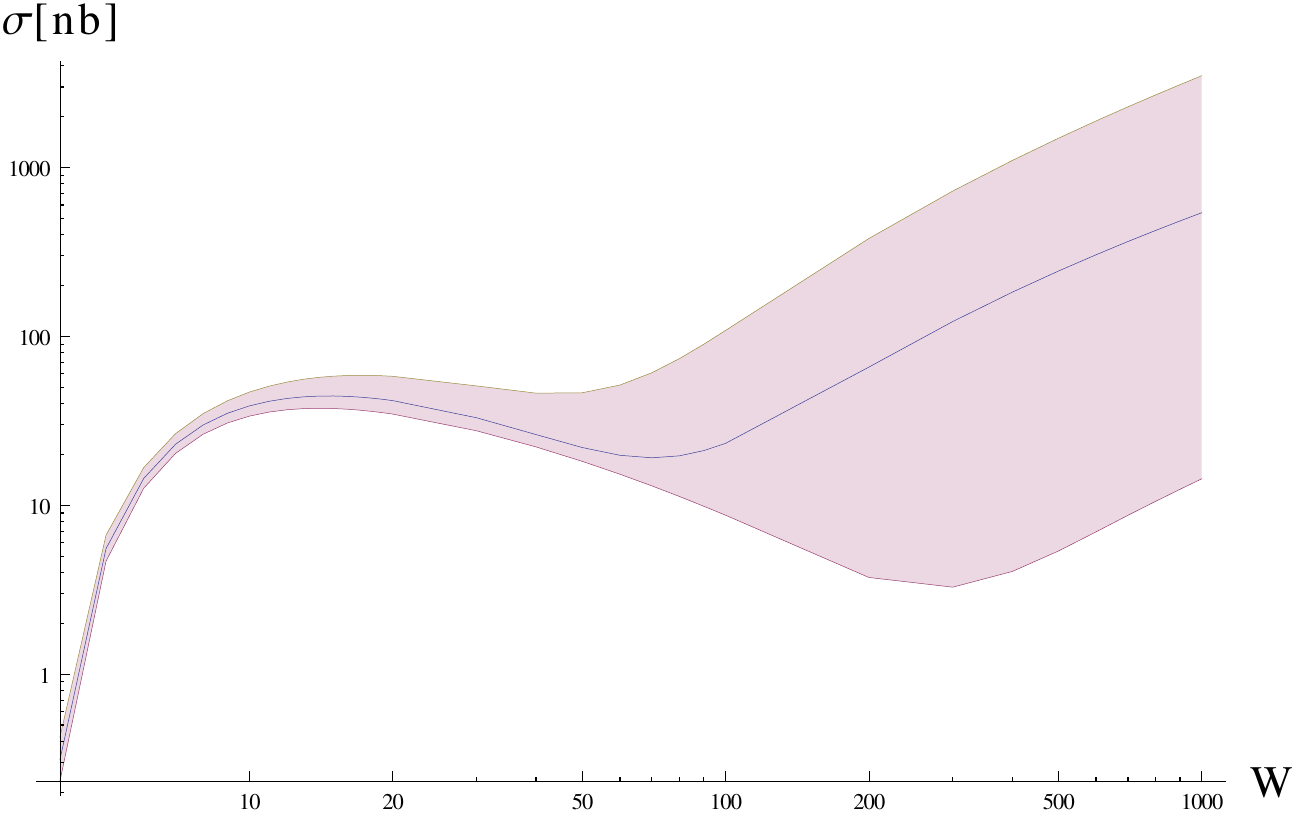}
\includegraphics[scale=.4]{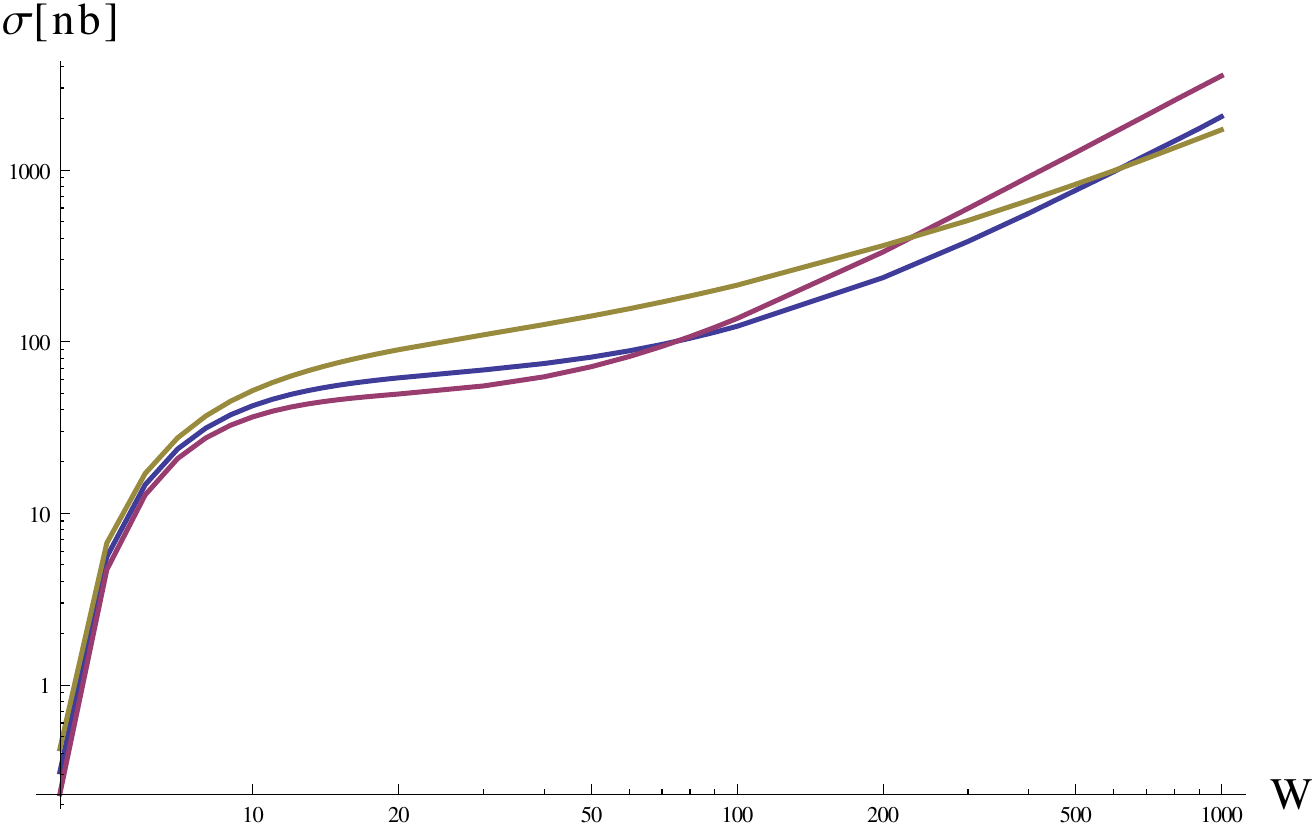}
\end{center}
\caption{NLO(left panel) and resummed (right panel) photoproduction cross section (only gluonic GPDs included in both cases) as a function of $W = \sqrt{s_{\gamma p}}$ for $\mu_F^2 = M_{J/\psi}^2 \times \{0.5,1,2\}$ (pink, blue and yellow lines respectively)}
\label{fig:resum_muF}
\end{figure}

Let us discuss  the value of resummed coefficient and its  dependence  on the scale of factorization $\mu_F$. Below we present results for two cases.
a) $\mu_F$ is equal to kinematic hard scale $\mu_F=M_V$ ($L=0$), and b)  $\mu_F$ is chosen to vanish the value of the first high energy term $C_1$, it requires $\mu_F=M_V/2=m$ ($L=\ln 4$):
\begin{eqnarray}
&
a) \ \ \ (\mu_F=M_V): \quad   1 -1.39\, z +2.61\, z^2 +0.481\, z^3 -4.96\, z^4 +\dots
& \nonumber \\
&
b) \ \ \ (\mu_F=M_V/2): \quad  1 +0.\, z +1.64\, z^2 +3.21\, z^3 +1.08\, z^4 +\dots \, .
& \nonumber
\end{eqnarray}
We see that almost all the high energy term coefficients, $C_n(L)$, have large absolute values. It shows that it is important
to take into account not only large NLO effects but also contributions from still higher orders of QCD collinear expansion that are enhanced
by the powers of large logarithms of energy. Another important observation is that it is not possible by appropriate choice of
factorization scale $\mu_F$ to move all enhanced by powers of $\ln(1/\xi)$ contributions from the coefficient function into the GPD (through
its $\mu_F$- evolution). Such a strategy is promoted in \cite{JMRT}. We see that our results above, the  case b), do not support this suggestion.
The choice $\mu_F=M_V/2$ indeed eliminates the big part of NLO correction from the hard coefficient, but it can not allow to get rid of such big terms
from the higher orders contributions. On the other hand, as illustrated on the Fig.\ref{fig:resum_muF}, the resummation up to the 6th order of the gluon GPD dependent coefficient function, greatly reduces the factorization scale dependence \footnote{In both cases (NLO and resummed) only the forward evolution of the PDFs, which enter the double distribution model for GPDs is performed.}. 

We believe that high energy resummation described above have to be incorporated in the analysis of
the  $J/\Psi$ and $\Upsilon$ meson photoproduction processes, this work is now in progress.












\section{Summary}
GDPs enter factorized amplitudes  for hard exclusive reactions  in a similar manner as PDFs enter factorized cross sections for inclusive DIS.
Ultraperipheral collisions at hadron colliders open a new way to measure GPDs in  TCS and photoproduction of heavy vector mesons at very small skewness parameters.
NLO corrections turn out to be rather large for TCS and even more
 for  vector meson  photoproduction  in the kinematics typical for experiments at the EIC collider. Various resummation techniques have been invoked to help to stabilize the perturbative expansion. 
In the heavy meson production case, high energy resummation techniques are suggested as a tool to provide reliable theoretical predictions in this kinematical domain.

\section{Acknowledgments}
This work is partly supported by the COPIN-IN2P3 Agreement, French grant ANR PARTONS (Grant No. ANR-
12-MONU-0008-01) and by the grant RFBR-15-02-05868,  L.Sz. was partially supported by grant of National Science Center, Poland, No. 2015/17/B/ST2/01838.

\end{document}